\documentclass{emulateapj}

\usepackage{amsmath}
\usepackage[sort&compress]{natbib}

\newcommand{\about}{$\sim\!\!$~}

\newcommand{\be}{\begin{displaymath}}
\newcommand{\ee}{\end{displaymath}}

\def\lsim{\hbox{\rlap{\raise 0.425ex\hbox{$<$}}\lower 0.65ex\hbox{$\sim$}}}
\def\gsim{\hbox{\rlap{\raise 0.425ex\hbox{$>$}}\lower 0.65ex\hbox{$\sim$}}}

\newcommand{\msun}{M$_\sun$}

\newcommand{\hbeta}{H$\beta$}
\newcommand{\hgamma}{H$\gamma$}
\newcommand{\hdelta}{H$\delta$}
\newcommand{\hepsilon}{H$\epsilon$}

\newcommand{\kms}{km~s$^{-1}$}
\newcommand{\ergps}{erg~s$^{-1}$}

\shorttitle{The Most Massive Stellar-Mass Black Hole}
\shortauthors{Silverman \& Filippenko}

\begin{document}

\title{On IC~10~X-1, the Most Massive Known Stellar-Mass Black Hole}

\author{Jeffrey M. Silverman and Alexei V. Filippenko}

\affil{Department of Astronomy, University of California,
Berkeley, CA 94720-3411; {jsilverman, alex}@astro.berkeley.edu}

\begin{abstract} 

IC~10~X-1 is a variable X-ray source in the Local Group starburst
galaxy \object{IC~10} whose optical counterpart is a Wolf-Rayet (WR)
star.  \citet{Prestwich07} recently proposed that it contains the most
massive known stellar-mass black hole (23-34 \msun), but their
conclusion was based on radial velocities derived from only a few
optical spectra, the most important of which was seriously affected by
a CCD defect. Here we present new spectra of the WR star, spanning one
month, obtained with the Keck-I 10 m telescope.  The spectra show a
periodic shift in the \ion{He}{2} $\lambda4686$ emission line as
compared with \object{IC~10} nebular lines such as [\ion{O}{3}]
$\lambda5007$.  From this, we calculate a period of $34.93\pm0.04$ hr
\citep[consistent with the X-ray period of $34.40\pm0.83$ hr reported by][]
{Prestwich07} and a radial-velocity semi-amplitude of $370\pm20$
\kms. The resulting mass function is $7.64\pm1.26$ \msun, consistent
with that of \citet{Prestwich07} (7.8 \msun). This, combined with the
previously estimated (from spectra) mass of 35 \msun\ for the WR star,
yields a minimum primary mass of $32.7\pm2.6$ \msun.  Even if the WR
star has a mass of only 17 \msun, the minimum primary mass is
$23.1\pm2.1$ \msun. Thus, IC~10~X-1 is indeed a WR/black-hole binary
containing the most massive known stellar-mass black hole.

\end{abstract}



\section{Introduction}\label{s:intro}

IC~10~X-1 is a bright, variable X-ray source in the Local Group
metal-poor starburst galaxy \object{IC~10} with an X-ray luminosity of
$10^{38}$ \ergps\ \citep{Brandt97,Bauer04}.  \citet{Lozinskaya07} suggested 
that IC~10~X-1 is the compact remnant of a hypernova, based on the nature of
the synchrotron superbubble in \object{IC~10}. There are four possible
optical counterparts to the X-ray source, the most likely being the
luminous Wolf-Rayet (WR) star [MAC92]~17A \citep{Crowther03}. Previous
spectroscopic observations of [MAC92]~17A show prominent \ion{He}{2}
line emission; \citet{Clark04} classified it as a WNE star.

\citet{Prestwich07} recently proposed that IC~10~X-1 and [MAC92]~17A
are a WR/black-hole (BH) binary containing the most massive known
stellar-mass black hole (23--34 \msun). Unfortunately, their
conclusion was based on radial velocity measurements from only a few
optical spectra, and the observation showing an apparent spectral
shift was seriously affected by a CCD defect, casting some doubt on
the reality of the shift.  They also assumed that the
observed X-ray period of IC~10~X-1 ($34.40\pm0.83$ hr) is equal to
the orbital period of the binary system.

Here we present radial velocities from 10 new optical 
spectra of the WR star spanning one month; our results confirm 
the conclusions of \citet{Prestwich07}.
In \S\ref{s:Observations} we describe the observations and data
reduction, in \S\ref{s:Analysis} we discuss our analysis of the
spectra, and in \S\ref{s:Results} we present our results.
A preliminary report on this work is given by \citet{Silverman08}.


\section{Observations and Data Reduction}\label{s:Observations}

[MAC92]~17A was observed with the Low Resolution Imaging Spectrometer
\citep[LRIS;][]{Oke95} mounted on the Keck-I 10~m telescope in 2007
during the nights of Nov.\ 11--13 (UT dates are used throughout this
paper), Nov.\ 16, and Dec.\ 12; a journal of observations is given in
Table~\ref{t:obs}. LRIS is now equipped with an atmospheric dispersion
compensator; thus, differential light losses \citep{Filippenko82} were 
not a problem, even at high airmasses.

\begin{deluxetable*}{cccccccc}
\tablecaption{Journal of Observations and Radial Velocities\label{t:obs}}
\tablewidth{0pt}
\tablehead{
\colhead{HJD\tablenotemark{a}} &
\colhead{2007 UT Date\tablenotemark{b}} &
\colhead{Exp. (s)} &
\colhead{Airmass\tablenotemark{c}} &
\colhead{Seeing\tablenotemark{d}} &
\colhead{Phase\tablenotemark{e}} &
\colhead{\ion{He}{2} $v$ (\kms)\tablenotemark{f}} &
\colhead{\ion{He}{2} FWHM (\AA)}}
\startdata
2454415.865 & Nov. 11.361 & 1400 & 1.33 & 0.8 & 0.9801 &  \phantom{$-$}$384\pm29$\phn\phn & $14.3\pm0.9$\phn \\
2454415.882 & Nov. 11.378 & 1400 & 1.36 & 0.8 & 0.9918 &  \phantom{$-$}$362\pm24$\phn\phn & $13.3\pm0.8$\phn \\
2454415.996 & Nov. 11.492 & 1800 & 1.95 & 1.5 & 0.0701 &  \phantom{$-$}$325\pm40$\phn\phn & $14.1\pm1.5$\phn \\
2454416.761 & Nov. 12.257 & 1800 & 1.35 & 0.7 & 0.5957 & $-312\pm44$\phn\phn & $18.1\pm1.7$\phn \\
2454416.932 & Nov. 12.428 & 1800 & 1.53 & 0.8 & 0.7132 &  \phn$-85\pm43$\phn\phn & $16.3\pm1.7$\phn \\
2454417.824 & Nov. 13.320 & 1800 & 1.30 & 1.5 & 0.3260 & $-182\pm26$\phn\phn & $15.5\pm0.9$\phn \\
2454420.708 & Nov. 16.205 & 1500 & 1.46 & 1.0 & 0.3081 & \phn$-77\pm113$\phn & $19.7\pm4.6$\phn \\
2454420.726 & Nov. 16.223 & 1500 & 1.40 & 1.1 & 0.3205 &$ -148\pm52$\phn\phn & $13.6\pm2.0$\phn \\
2454420.745 & Nov. 16.242\tablenotemark{g} & 1500 & 1.36 & 1.0 & 0.3335 & $-133\pm51$\phn\phn & $14.0\pm1.9$\phn \\
2454446.761 & Dec. 12.258 & 1200 & 1.31 & 1.4 & 0.2074 &   \phantom{$-$}\phn$91\pm51$\phn\phn & $24.6\pm2.1$\phn
\enddata
\tablenotetext{a}{Heliocentric Julian Date at midpoint of exposure.}
\tablenotetext{b}{UT Date at midpoint of exposure.}
\tablenotetext{c}{Airmass at midpoint of exposure.}
\tablenotetext{d}{Approximate full width at half-maximum intensity (FWHM), arcsec.}
\tablenotetext{e}{Using $P = 34.93\pm0.04$ hr and $T_0$ = 2007 Nov.\ $9.935\pm0.014$
(the UT date of maximum radial velocity).}
\tablenotetext{f}{Radial velocity calculated relative to the [\ion{O}{3}]
$\lambda5007$ line.}
\tablenotetext{g}{This is the only observation with a long slit of width 1\farcs0;
all others used a slit of width 0\farcs7.}
\end{deluxetable*}

Fortuitously, the long slit included RSMV~2, another WR star in
\object{IC~10} \citep{Crowther03}.  Unfortunately, the light from this second
object fell between the two chips on the blue side of LRIS in the
observations on Nov. 11.492 and Nov. 13.320.

All data were taken using a 5600~\AA\ dichroic and a 600/4000 grism on
the blue side of the spectrograph. 
All but one observation employed a long slit of width 0\farcs7; the
observation on 2007 Nov. 16.242 used a long slit of width 1\farcs0.
The slit was always oriented at a position angle (PA) of 162$^\circ$,
through another star readily visible on the acquisition and guider
image, in order to facilitate object acquisition and to minimize
contamination from adjacent stars. The spectra span a wavelength range
of \about3800--5550~\AA, and all observations had a FWHM spectral
resolution
of \about3.5~\AA\ (\about220 \kms).

The data were reduced using standard techniques
\citep[e.g.,][]{Foley03}.  Routine CCD processing and spectrum
extraction for the data were completed with IRAF\footnote{IRAF, the
Image Reduction and Analysis Facility, is distributed by the National
Optical Astronomy Observatories, which is operated by the Association
of Universities for Research in Astronomy, Inc. (AURA) under
cooperative agreement with the National Science Foundation (NSF).}.
The data were extracted with the optimal algorithm of
\citet{Horne86}. We obtained the overall wavelength scale from
low-order polynomial fits to calibration-lamp spectra.  Small
wavelength shifts were applied to individual spectra after
cross-cor\-re\-lat\-ing night-sky lines with a template sky spectrum.
Using our own IDL routines, we fit spectrophotometric standard-star
spectra to flux-calibrate our data and remove telluric lines
\citep{Wade88, Matheson00}.

\begin{figure}
\epsscale{0.8}
\rotatebox{90}{
\plotone{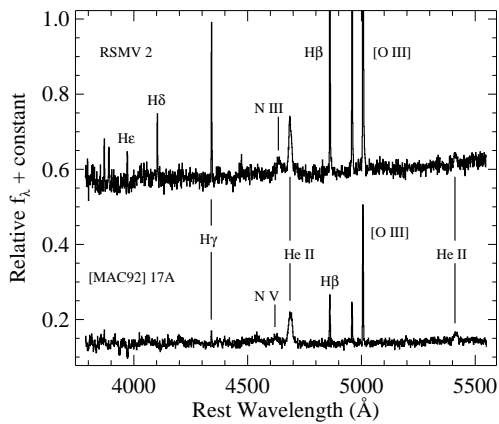}}
\caption{Unphased combined spectra of RSMV~2 ({\it top}) and [MAC92]~17A 
({\it bottom}), and their superposed \ion{H}{2} regions. Both nebular 
and WR line identifications are discussed in the text. The spectrum of
RSMV~2 has been shifted up by 0.45 units for clarity.}
\label{f:bothcombspec}
\end{figure}

The unphased combined spectra of RSMV~2, from 8 epoches, and
of [MAC92]~17A, from all 10 epochs, are
shown in Figure~\ref{f:bothcombspec}.
The radial velocity of \object{IC~10}, $-348\pm1$ \kms\ \citep{Huchra99}, has
been removed.

In the spectra of both objects, there are
strong nebular emission lines of \hgamma, \hbeta, and [\ion{O}{3}]
(and \hdelta\ and \hepsilon\ in RSVM~2) from
\ion{H}{2} regions in the slit. One can also see the
relatively broad \ion{He}{2} $\lambda4686$ emission and the weaker
\ion{He}{2} $\lambda5411$ emission which comes from
the WR stars themselves.  In addition, the fainter \ion{N}{5}
4603--20~\AA\ and \ion{N}{3} 4634--42~\AA\ emission features
from [MAC92]~17A and RWVM~2 (respectively) are seen.
Broad emission from relatively low-excitation species such as \ion{H}{1}
and \ion{He}{1} is not present in [MAC92]~17A, suggesting that the
emission does not arise from an accretion disk.


\section{Analysis}\label{s:Analysis}

The X-ray period found by \citet{Prestwich07}, as well as our
measured period (see \S\ref{ss:rvmf}), are substantially larger
than the
period needed for Roche lobe overflow, given reasonable masses of
[MAC92]~17A and its compact binary companion.
(We, however, derive an expected period for Roche lobe overflow
of 9.0--15.9 hr, which
differs from the 2--3.5 hr period suggested by Prestwich et al.\ 2007).
Thus, the accretion
must be the result of a wind from the WR
star, as is often seen in high-mass X-ray binaries. Furthermore, the
\ion{He}{2} $\lambda4686$ emission line is formed in the inner part of
the WR wind (close to the star), justifying its use when
determining radial velocities of WR binaries \citep{Prestwich07}.

For each observation we fit a Gaussian profile to the nebular
[\ion{O}{3}] and \hbeta~lines, as well as to the \ion{He}{2} line of
[MAC92]~17A. To remove possible errors introduced by our wavelength
and flux calibrations, the profiles were fit to the raw,
one-dimensional (1-D) extracted spectra (i.e., neither wavelength nor
flux calibrated). Owing to the small number of emission lines available
in the calibration-lamp spectra, the wavelength solution was found to 
vary quite a bit from spectrum to spectrum, when in fact nothing was
actually changing except perhaps the zero point.
Fitting profiles to this raw
1-D data is reasonable for our purposes because the
radial velocity of the WR star only depends on the {\it relative} change in
the spacing between the \ion{He}{2} line and the (stationary) nebular
lines. 

\begin{figure}
\epsscale{1.2}
\rotatebox{90}{
\plotone{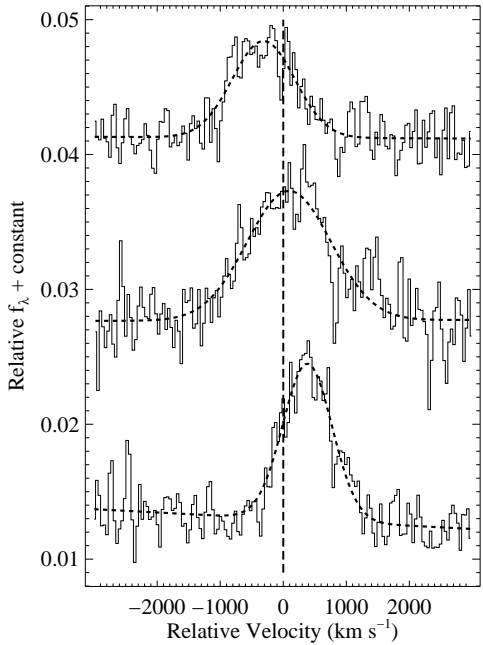}}
\caption{Spectra of the \ion{He}{2} emission line from
[MAC92]~17A obtained on 2007 Nov.\ 12.257 ({\it top}), 2007
Dec.\ 12.258 ({\it middle}), 
and 2007 Nov.\ 11.361 ({\it bottom}).  The data
have been deredshifted and converted into velocity
space.  The vertical long-dashed line is where the 
\ion{He}{2} feature of [MAC92]~17A would be
centered if at rest. The short-dashed lines show our
Gaussian fits to the data.  The centroid of the \ion{He}{2} 
line clearly shifts with time.}\label{f:three_epochs}
\end{figure}

Figure~\ref{f:three_epochs} shows three spectra, centered on the
\ion{He}{2} $\lambda4686$ spectral feature, in velocity space (as
calculated relative to the [\ion{O}{3}] $\lambda5007$ line).  The
signal-to-noise ratio (S/N) of the observations is typical of our
entire dataset.  It is clear that the centroid of the \ion{He}{2} line
is shifted between the three observations.

The FWHM of the \ion{He}{2} line appears to change with time
in Figure~\ref{f:three_epochs}. As seen in the last column of 
Table~\ref{t:obs}, the only observation that actually has a substantially
different FWHM is the one from 2007 Dec.\ 12.258 (the middle spectrum in
Figure~\ref{f:three_epochs}). \citet{Prestwich07} note that most of their
epochs were well-fit by a Gaussian with a 15--17~\AA\ FWHM, which matches
our data quite well (we have a mean of \about16.3~\AA).  They also
mention that one of their observations had a FWHM of only 10~\AA, but that
the narrowness of this observation is likely due to a ``CCD defect that
contaminates the red component.'' None of our derived FWHM values are
within 1$\sigma$ of 10~\AA, suggesting that their assessment is 
correct. However, this is also the only spectrum from which 
\citet{Prestwich07} deduced a radial velocity variation in [MAC92]~17A,
casting some doubt on the claimed shift.

From our Gaussian fits we calculated the separation in pixels between
the centroids of the \ion{He}{2} feature and each host-galaxy nebular 
line. These were then converted to a difference in wavelength by using 
the dispersion value we derived from low-order polynomial fits to
calibration-lamp spectra (0.621~\AA\ pixel$^{-1}$). Next, we performed a
nonlinear least-squares fit to the differences in wavelength. The
resulting cosine yielded the period of the \ion{He}{2} line and its
systemic wavelength shift, which is equivalent to the
zero point of the cosine. Finally, each spectrum's
deviation from this systemic wavelength shift was converted into a
radial velocity. Similarly, the semi-amplitude of the fit was
converted from a wavelength to a velocity.


\section{Results}\label{s:Results}

\subsection{Radial-Velocity Curve}\label{ss:rvmf}

\begin{figure}
\epsscale{0.75}
\rotatebox{90}{
\plotone{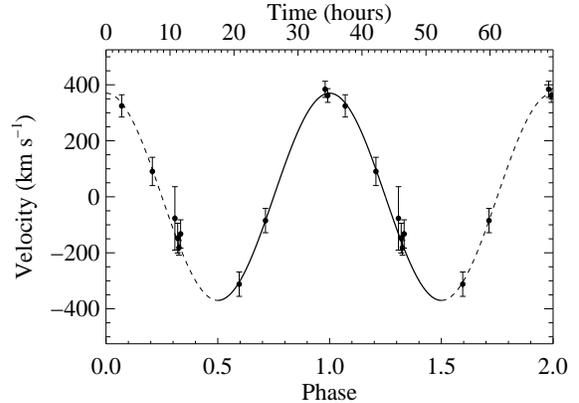}}
\caption{Radial-velocity curve of [MAC92]~17A using velocities relative to
the [\ion{O}{3}] $\lambda5007$ spectral line. Two cycles are shown for 
clarity. Formal velocity error bars are 1$\sigma$.
See the text for values of the fit parameters.}\label{f:rv}
\end{figure}

Figure~\ref{f:rv} shows the radial-velocity curve of [MAC92]~17A. Two
complete periods, each with 10 observations, are shown.  The
velocities in Figure~\ref{f:rv} were calculated with respect to the
[\ion{O}{3}] $\lambda5007$ line. We also determined radial velocities
and an associated cosine fit with respect to the nebular \hbeta\ line.
These radial velocities differed from the ones derived using the [\ion{O}{3}]
line by much less than $1\sigma$ at each epoch.
In addition, the period of the curve derived using the \hbeta\ line
differed by only 0.01\% from the
curve shown in Figure~\ref{f:rv}, and the semi-amplitude differed by \about0.3\%.
These differences led to a mere 0.5\% difference in the final BH mass.

Since [\ion{O}{3}] had a somewhat higher S/N than \hbeta\ (as seen in
Figure~\ref{f:bothcombspec}), the \hbeta\ Gaussian fits had slightly
larger uncertainties in their centroids and thus led to larger overall
uncertainties. Thus, we have chosen to present only the velocities
derived relative to the [\ion{O}{3}] line.

The error bars shown represent $1\sigma$, and were
calculated by considering the uncertainty in the centroids of our
Gaussian fits to both the \ion{He}{2} line as well as the [\ion{O}{3}]
nebular line.  The point in Figure~\ref{f:rv} with the relatively
large error bar is due to a low S/N spectrum.

The four-parameter fit (zero point, semi-am\-pli\-tude, period, and
phase) yielded a semi-amplitude of $K_2 = 370\pm20$ \kms\ and a period
of $P = 34.93\pm0.04$ hr (consistent with the X-ray period of
$34.40\pm0.83$ hr reported by \citealt{Prestwich07}). However, the
phasing of our radial-velocity curve with the X-ray light curve
\citep[][Fig. 1]{Prestwich07} is unknown because of the substantial
uncertainty in the X-ray period.

\begin{figure}
\epsscale{0.65}
\rotatebox{90}{
\plotone{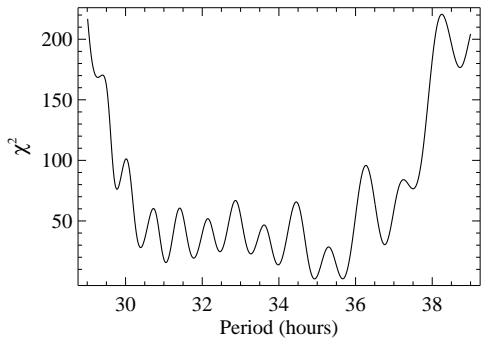}}
\caption{Value of $\chi^2$ versus trial period for [MAC92]~17A, using all 
10 spectra discussed in the text.}\label{f:chisquared}
\end{figure}

We examined the distribution of possible periods
resulting from our observations in order to test whether our
calculated period suffers from any ambiguity in the number of cycles
between observations. Figure~\ref{f:chisquared} shows $\chi^2$
versus trial period for our complete set of 10
observations.  It is clear that the two periods that best fit the data
are \about34.9 hr ($\chi^2 \approx 2.43$) and \about35.7 hr
($\chi^2 \approx 2.50$).

We have chosen to adopt the 34.9 hr period since (a) it
has the lowest value of $\chi^2$ (although the $\chi^2$ from
the 35.7 hr period is close to this value); (b) it
is consistent with the X-ray period derived by \citet{Prestwich07},
whereas the 35.7 hr period is 2$\sigma$ away from this value; and (c) it
is surrounded by the lowest values of $\chi^2$ in
Figure~\ref{f:chisquared} (yet the value of $\chi^2$ begins to rise
dramatically at periods just larger than 35.7 hr).
The fact that we cannot completely discern between the two periods is the
result of a half-cycle ambiguity between our Nov.\ and Dec.\ observations.
Clouds precluded a second exposure on Dec.\ 12 which would have almost
certainly eliminated this uncertainty.

\subsection{Determination of the Black Hole Mass}\label{ss:BHmass}

Given a radial-velocity curve of the secondary star, two parameters are 
necessary to calculate the minimum mass of the companion BH:
the period ($P$) and the semi-amplitude ($K_2$).  The standard method of 
measuring stellar masses in binary systems is to calculate the mass 
function, $f\left(M\right)$ \citep[e.g., ][]{McClintock06},
\begin{equation}
f\left(M_1\right) \equiv \frac{PK_2^3}{2\pi G} = \frac{M_1\sin^3i}{\left(1+q\right)^2}
\label{e:massfunc},
\end{equation}
where $i$ is the inclination of the binary's orbit and $q=M_2/M_1$;
here, $M_1$ is the mass of the putative BH and $M_2$ is the mass of
the WR donor. Physically, $f\left(M_1\right)$ is the smallest possible 
BH mass; it is equal to the BH mass only if $i = 90^\circ$ and $q = 0$.

Using our measured values of $P$ and $K_2$ (see \S\ref{ss:rvmf} above),
we calculated the mass function for the system.  Finally, we assumed a
range of reasonable masses for the WR star, which then allowed us to
calculate a range of possible BH masses.

Adopting our calculated period and semi-amplitude, the resulting mass
function using Equation~\ref{e:massfunc} is $7.64\pm1.26$ \msun,
consistent with that of \citet{Prestwich07} ($7.8$ \msun).  If we
assume a mass for the WR companion and an inclination for the orbit,
then we can calculate a mass for the BH.

The spectroscopic mass of the WR star was calculated by
\citet{Clark04} to be 35 \msun, but there is much uncertainty in this
value. \citet{Prestwich07} point out the small possibility that
[MAC92]~17A could have a mass as low as 17 \msun. They also note that
since this system is observed to have deep X-ray eclipses, the
inclination of the system should be close to 90$^\circ$.

Table~\ref{t:masses} shows possible BH masses assuming a reasonable
range of values for both the mass of [MAC92]~17A and the orbital
inclination \citep[similar to][Table 1]{Prestwich07}.  Given these
ranges and our calculated values for the period and semi-amplitude of
the radial-velocity curve, the BH in IC~10~X-1 must have a mass of
\textit{at least} 23.1 \msun, and it very well might be 32.7 \msun\ or
larger. In any case, this is much larger than the mass of the BH in M33,
$15.65\pm1.45$ \msun\ \citep{Orosz07}, arguably the most massive
previously identified stellar-mass BH, which was announced very shortly 
before the BH mass of IC~10~X-1 \citep{Prestwich07}.

Note that if we had used the period of 35.7 hr mentioned in \S\ref{ss:rvmf}
and its associated
semi-amplitude of 390 \kms, we would have derived a mass function of
$9.15\pm1.45$ \msun, yielding minimum BH masses of \about25.5--35.8
\msun\ (given the previously assumed range of WR masses, 17--35 \msun).

\begin{deluxetable}{cccc}
\tablecaption{Derived Black Hole Mass (\msun)\label{t:masses}}
\tablewidth{0pt}
\tablehead{
\colhead{Inclination} &
\multicolumn{3}{c}{Wolf-Rayet Mass (\msun)} \\
\cline{2-4} \\[-7pt]
\colhead{(deg)} &
\colhead{17} &
\colhead{25} &
\colhead{35}}
\startdata
90 & $23.1\pm2.1$\phn & $27.7\pm2.3$\phn & $32.7\pm2.6$\phn \\
78 & $23.9\pm2.1$\phn & $28.6\pm2.4$\phn & $33.8\pm2.8$\phn \\
65\tablenotemark{a} & $27.1\pm2.5$\phn & $32.3\pm2.8$\phn & $37.9\pm3.2$\phn
\enddata
\tablenotetext{a}{If the mass of [MAC92]~17A is 35 \msun, inclinations less than
\about78$^\circ$ will not yield X-ray eclipses.}
\end{deluxetable}

\section{Conclusion}\label{s:conclusion}

In this paper, we present new spectra of the WR stars [MAC92]~17A and
RSMV~2 in the nearby starburst galaxy \object{IC~10}.  [MAC92]~17A has
been shown to be in a binary system with the variable X-ray source
IC~10~X-1 \citep{Prestwich07}.  From our Keck spectra of [MAC92]~17A, we
have constructed a compelling radial-velocity curve. The measured
orbital period and semi-amplitude of [MAC92]~17A imply that the
mass of the BH companion is at least 23.1 \msun, and more likely
$\sim$32.7 \msun. Thus, we have shown that IC~10~X-1 is indeed a WR/BH
binary containing the most massive known stellar-mass BH.

\acknowledgements
The data presented herein were obtained at the W. M. Keck Observatory,
which is operated as a scientific partnership among the California
Institute of Technology, the University of California, and the
National Aeronautics and Space Administration.  The Observatory was
made possible by the generous financial support of the W. M. Keck
Foundation. We wish to recognize and acknowledge the very significant
cultural role and reverence that the summit of Mauna Kea has always
had within the indigenous Hawaiian community; we are most fortunate
to have the opportunity to conduct observations from this mountain. 
We thank M. Bogosavljevic, C. C. Steidel, D. J. Neill, and M. Seibert for
obtaining some of the Keck spectra on our behalf. We are also grateful
to R. J. Foley and R. Chornock for helpful discussions. This work was
supported by the NSF through grant AST--0607485.



\begin{thebibliography}{16}
\expandafter\ifx\csname natexlab\endcsname\relax\def\natexlab#1{#1}\fi

\bibitem[{{Bauer} \& {Brandt}(2004)}]{Bauer04}
{Bauer}, F.~E., \& {Brandt}, W.~N. 2004, \apjl, 601, L67

\bibitem[{{Brandt} {et~al.}(1997){Brandt}, {Ward}, {Fabian}, \&
  {Hodge}}]{Brandt97}
{Brandt}, W.~N., {Ward}, M.~J., {Fabian}, A.~C., \& {Hodge}, P.~W. 1997,
  \mnras, 291, 709

\bibitem[{{Clark} \& {Crowther}(2004)}]{Clark04}
{Clark}, J.~S., \& {Crowther}, P.~A. 2004, \aap, 414, L45

\bibitem[{{Crowther} {et~al.}(2003){Crowther}, {Drissen}, {Abbott}, {Royer}, \&
  {Smartt}}]{Crowther03}
{Crowther}, P.~A., {Drissen}, L., {Abbott}, J.~B., {Royer}, P., \& {Smartt},
  S.~J. 2003, \aap, 404, 483

\bibitem[{{Filippenko}(1982)}]{Filippenko82}
{Filippenko}, A.~V. 1982, \pasp, 94, 715

\bibitem[{{Foley} {et~al.}(2003){Foley}, {Papenkova}, {Swift}, {Filippenko},
  {Li}, {Mazzali}, {Chornock}, {Leonard}, \& {Van Dyk}}]{Foley03}
{Foley}, R.~J., {Papenkova}, M.~S., {Swift}, B.~J., {Filippenko}, A.~V., {Li},
  W., {Mazzali}, P.~A., {Chornock}, R., {Leonard}, D.~C., \& {Van Dyk}, S.~D.
  2003, \pasp, 115, 1220

\bibitem[{{Horne}(1986)}]{Horne86}
{Horne}, K. 1986, \pasp, 98, 609

\bibitem[{{Huchra} {et~al.}(1999){Huchra}, {Vogeley}, \& {Geller}}]{Huchra99}
{Huchra}, J.~P., {Vogeley}, M.~S., \& {Geller}, M.~J. 1999, \apjs, 121, 287

\bibitem[{{Lozinskaya} \& {Moiseev}(2007)}]{Lozinskaya07}
{Lozinskaya}, T.~A., \& {Moiseev}, A.~V. 2007, \mnras, 381, L26

\bibitem[{{Matheson} {et~al.}(2000){Matheson}, {Filippenko}, {Ho}, {Barth}, \&
  {Leonard}}]{Matheson00}
{Matheson}, T., {Filippenko}, A.~V., {Ho}, L.~C., {Barth}, A.~J., \& {Leonard},
  D.~C. 2000, \aj, 120, 1499

\bibitem[{{McClintock} \& {Remillard}(2006)}]{McClintock06}
{McClintock}, J.~E., \& {Remillard}, R.~A. 2006, in {Compact Stellar X-Ray
  Sources}, ed. W.~{Lewin} \& M.~{van der Klis} (Cambridge: Cambridge
  University Press), 157

\bibitem[{{Oke} {et~al.}(1995){Oke}, {Cohen}, {Carr}, {Cromer}, {Dingizian},
  {Harris}, {Labrecque}, {Lucinio}, {Schaal}, {Epps}, \& {Miller}}]{Oke95}
{Oke}, J.~B., {Cohen}, J.~G., {Carr}, M., {Cromer}, J., {Dingizian}, A.,
  {Harris}, F.~H., {Labrecque}, S., {Lucinio}, R., {Schaal}, W., {Epps}, H., \&
  {Miller}, J. 1995, \pasp, 107, 375

\bibitem[{{Orosz} {et~al.}(2007){Orosz}, {McClintock}, {Narayan}, {Bailyn},
  {Hartman}, {Macri}, {Liu}, {Pietsch}, {Remillard}, {Shporer}, \&
  {Mazeh}}]{Orosz07}
{Orosz}, J.~A., {McClintock}, J.~E., {Narayan}, R., {Bailyn}, C.~D., {Hartman},
  J.~D., {Macri}, L., {Liu}, J., {Pietsch}, W., {Remillard}, R.~A., {Shporer},
  A., \& {Mazeh}, T. 2007, \nat, 449, 872

\bibitem[{{Prestwich} {et~al.}(2007){Prestwich}, {Kilgard}, {Crowther},
  {Carpano}, {Pollock}, {Zezas}, {Saar}, {Roberts}, \& {Ward}}]{Prestwich07}
{Prestwich}, A.~H., {Kilgard}, R., {Crowther}, P.~A., {Carpano}, S., {Pollock},
  A.~M.~T., {Zezas}, A., {Saar}, S.~H., {Roberts}, T.~P., \& {Ward}, M.~J.
  2007, \apjl, 669, L21

\bibitem[{{Silverman} \& {Filippenko}(2008)}]{Silverman08}
{Silverman}, J., \& {Filippenko}, A.~V. 2008, American Astronomical Society
  Meeting Abstracts, Vol. 211, 161.06

\bibitem[{{Wade} \& {Horne}(1988)}]{Wade88}
{Wade}, R.~A., \& {Horne}, K. 1988, \apj, 324, 411

\end{thebibliography}

\end{document}